\documentclass[12pt,showpacs,preprint]{revtex4}
\usepackage{pstricks,pst-node}         
\usepackage[dvips]{graphicx}           
\usepackage{xspace}                    
\usepackage{amssymb,latexsym,amsmath}  
\usepackage{setspace}                  
\usepackage{layout}
\newcommand{\locus}[1][a]{\ensuremath{\mathbf{S}^{#1}}\xspace}
\newcommand{\locusn}{\ensuremath{\mathbf{S}}\xspace}
\newcommand{\fit}{\ensuremath{p_{off}}\xspace}
\newcommand{\kil}{\ensuremath{p_{kill}}\xspace}
\newcommand{\mut}{\ensuremath{p_{mut}}\xspace}
\newcommand{\ert}{\ensuremath{p_{mut}^{th}}\xspace}

\newpsobject{showgrid}{psgrid}{subgriddiv=1,griddots=10,gridlabels=6pt}
\begin{document}

\title{The Tangled Nature model as an evolving quasi-species model}
\author{Simone Avogadro di Collobiano}
\affiliation{Blackett Laboratory, Imperial College London, Prince Consort Road, London SW7 2BW, United Kingdom}
\author{Kim Christensen}
\email{k.christensen@imperial.ac.uk}
\homepage{http://www.cmth.ph.imperial.ac.uk/kim/}
\affiliation{Blackett Laboratory, Imperial College London, Prince Consort Road, London SW7 2BW, United Kingdom}
\author{Henrik Jeldtoft Jensen}
\email{h.jensen@imperial.ac.uk}
\homepage{http://www.ma.imperial.ac.uk/~hjjens/}
\affiliation{Department of Mathematics, Imperial College London, 180 Queen's Gate, London SW7 2BZ, United Kingdom}

\begin{abstract}
We show that the Tangled Nature model can be interpreted as a general
formulation of the quasi-species model by Eigen {\it et al.} in a frequency
dependent fitness landscape. We present a detailed theoretical derivation
of the mutation threshold, consistent with the simulation results,
that provides a valuable insight into how the microscopic dynamics of the model
determine the observed macroscopic phenomena published previously.
The dynamics of the Tangled Nature model is defined
on the microevolutionary time scale via reproduction, with heredity,
variation, and natural selection.
Each organism reproduces with a rate that is linked to the individuals'
genetic sequence and depends on the composition of the population
in genotype space. Thus the microevolutionary dynamics of the fitness
landscape is regulated by, and regulates, the evolution of the
species by means of the mutual interactions.
At low mutation rate, the macro evolutionary pattern mimics the fossil data:
periods of stasis, where the population is concentrated in a network of
coexisting species, is interrupted by bursts of activity. As the
mutation rate increases, the duration and the frequency of bursts increases.
Eventually, when the mutation rate reaches a certain threshold, the
population is spread evenly throughout the genotype space showing that
natural selection only leads to multiple distinct species if adaptation is
allowed time to cause fixation.
\end{abstract}


\maketitle

\section{Introduction}
\label{sec:intro}
 Explaining observed macro-evolutionary patterns as collective emergent
properties of systems with many interacting degrees of freedom, whether
these be single individuals or `species', is an alluring challenge for
researchers with a background in statistical
physics~\cite{car:big,Pel:staDar}.
The quasi-species model by Eigen {\it et al.} \cite{eig:sel,eig:qua}
has proved useful when investigating the behaviour of populations
in a given fixed fitness landscape and it provides a firm paradigm for many
models~\cite{dro:bio}.

The fundamental idea in the approach by Eigen {\it et al.} is
to identify species with sequences in genotype space. Positions
in genotype space which are assigned particularly high fitness are
called wildtypes, that is, the forms that predominate
in a population are well adapted to the environment.  During the
reproduction event, mutations are seen as errors of the replication of
the parental sequence. The effect is thus to spread the population
from the original point to neighbouring positions in genotype space. If
one were to use a classical Darwinian view on such a process, the
population would then be sharply localised in genotype space on
the position which corresponds to a high fitness: all other positions
would be cancelled by their low fitness. This can only be true if the
replication process is by and large accurate. A replication process with a
too high mutation rate would produce copies of the original
fit parent with so many errors that selection is unable to maintain the
population at the original point.

By lowering the mutation rate progressively, variation would be less
effective in dispersing the population since the offspring are more
similar to the parents.
The quasi-species model defines the presence of a threshold, in
the mutation rate, where the multiplication process changes
drastically. The gradual decrease of the mutation rate sees the
transition from a random population, diffused as scattered points in
genotype space, to a population constrained to a few positions.

The transition of the process from a random state to an ordered one is
a phase transition, with the mutation rate acting as a control
parameter.  The nature of the transition has been extensively studied.
In the seminal paper by Eigen {\it et al.}, where the transition was first
noticed~\cite{eig:qua}, the species have a predetermined fixed fitness
associated.  Subsequently, the quasi-species model has been analysed in
different fitness landscapes~\cite{pel:err,pel:mea}, for different
topologies of the genotype space~\cite{kol:top}, and for spatially
resolved models~\cite{alt:err}, each confirming the original results.
Finally, the error threshold in a model with a
dynamical fitness landscape~\cite{nil:err} has been analysed.
In this case however, the dynamics is
regulated artificially from outside.

It has been shown that it is possible to map the original quasi-species
model onto a two-dimensional Ising system with nearest-neighbour
interaction in one direction~\cite{leu:exa}, and that, in this
representation, for simple fitness landscapes, the correspondence
links the error threshold with a first-order phase
transition~\cite{tar:err}.
A relation of fundamental importance by Galluccio
\textit{et al.}~\cite{gal:dif}  proves that the error-threshold naturally
arises as a consequence of the model introduced and that, more
generally, for a given mutation rate \mut and a given reproduction
rate \fit it is possible to determine uniquely an upper limit for the
length of the genetic sequence.

We show that the Tangled Nature model, introduced
in detail in Refs.~\cite{chr:tan, hal:agi}, can be considered as a
general formulation of the quasi-species model. The generalisation is
provided by a relaxation on the condition of fixed population size,
which, in the original formulation, acts as selection principle on the
sequences. The most important features of the Tangled Nature model,
for details see Refs. \cite{chr:tan,hal:agi},
is that of creating multiple co-evolving quasi-species in a
frequency dependent fitness landscape, where the dynamics of
the landscape is an inherent property of the model. 
In this paper, we present in detail the theoretical calculation of the mutation
threshold which fits the experimental accurately \cite{hal:agi}.

It is also interesting to point out the connection between the Tangled
Nature model and game theoretical non-linear replicator dynamics
\cite{Chap7}. In both cases the
reproduction of a given type of individuals depends on the configuration
of the entire population. One therefore expects to find stable solutions
to the dynamics of the Tangled Nature model similar to the Nash
equilibria or Evolutionary Stable Strategies found for replicator
dynamics. We have stressed this relation by using the term
quasi-Evolutionary Stable Strategies to denote
the quasi-stable configurations of the Tangled Nature model.

In Sec. \ref{sec:quasi} we review briefly the quasi-species model
by Eigen \textit{et al.} Section \ref{sec:TaNa} briefly discusses
the definition of the Tangled Nature model with an intrinsically
generated dynamic fitness landscape. We discuss in detail the dynamics
of the model in terms of difference equations.
Section \ref{sec:error} contains a discussion of the error threshold
theoretically and numerically and finally in Sec. {\ref{diss} we discuss
the relation between the Tangled Nature model and the
quasi-species model by Eigen {\it et al.}

\section{The quasi-species model}
\label{sec:quasi}
Eigen \textit{et al.}~\cite{eig:qua} introduced a model in which the
effects of various mutation rates on a process of replication of
finite sequences of binary values is explored. Each sequence 
$\locus = \{S^{a}_{1},S^{a}_{2},\ldots,S^{a}_{L} \}$, where $S_{i}^{a} =
\{-1,+1\}, \ \ i = 1,2,\ldots,L$ in genotype space represents
a species. Each existing sequence $\locus$ replicates, with a constant
rate $p_{off}^{a} = p_{off}(\locus)$ and degrades with a constant and
universal (i.e., independent of position) rate $p_{kill}$.
The number $n_{a}(t) = n(\locus)(t)$ of copies of a given sequence
\locus varies with time.
The replication process is not exact but prone to error. During the
replication, the rate of mutation per gene is \mut.

The model has been solved analytically in the limit where one
particular sequence is assumed to have a high fitness, while
mutants are less fit.
For low mutation rate, the population is concentrated
around the top of the mountain in the fitness landscape.
The dominant sequence with its surrounding mutants is called
a {\it quasi-species}.
As the mutation rate increases, the population drifts away from the top down
to the ridges. Eventually, when the mutation rate reaches a threshold
value $\ert$, the population is spread evenly throughout the fitness
landscape, that is, a phase transition occurs at $\ert$.

\section{The Dynamics of the Tangled Nature model}
\label{sec:TaNa}

The dynamics of the Tangled Nature model is defined via an elementary time
step where (a) one organism is randomly selected and killed with 
constant probability $p_{kill}$ (b) one organism
is randomly selected and with probability
$p_{off}$, that depends on the
current composition of the population in genotype space, two offspring
are reproduced and the parent is then removed from the ecology
\cite{chr:tan, hal:agi}.

By analysing the dynamics it is possible to characterise the stable
configurations that may develop in the Tangled Nature Model.

The difference equation describing the variation of the number of
individuals of a position \locus during a single time step can be
derived as follow.  Let $n_{a}(t)$ denote the number of individuals at
position \locus. Then
\begin{equation}
  n_{a}(t + 1) = n_{a}(t) + \sum_{E} \Delta n_{a}(E) \cdot P(E),
\end{equation}
where $t$ is the number of time steps, $E$ refers to any event that
can affect $n_{a}$, that is, an annihilation event or reproduction
event, by an amount $\Delta n_{a}(E)$. The event $E$ occurs with
probability $P(E)$.

For a killing event, $\Delta n_{a}(E) = -1$ and the probability of a
killing event is the product of the probability of choosing an
organism in position \locus times the killing rate, that is, $P(E) =
\rho_{a}(t)\kil$, where we have introduced the density 
$\rho_{a}(t) = \frac{n_{a}(t)}{\sum_a n_{a}(t)}$
of organisms at position \locus.

For a reproduction event, distinction has to be made between the case
where reproduction originates from position \locus, see Fig. \ref{fig1}(a)
and reproduction
originating from any other position \locus[b] different from \locus,
which we will call the ``back-flow'' contribution, see Fig. \ref{fig1}(b).

\begin{figure}[htpb]
\centering \begin{pspicture}(15,3.5)

  \rput(0.25,3.5){(a)}

 \rput(2.625,3.25){\small $E$}

 \pscircle(0.5,1.25){0.25} \psline(0,1.75)(1,0.75) \psline(0,0.75)(1,1.75)

 \pscircle(2.25,2.25){0.25} \pscircle(3,2.25){0.25}

 \pscircle[fillstyle=solid,fillcolor=black](2.25,1.25){0.25}
 \pscircle(3,1.25){0.25}

 \pscircle[fillstyle=solid,fillcolor=black](2.25,0.25){0.25}
 \pscircle[fillstyle=solid,fillcolor=black](3,0.25){0.25}

 \rput{26.57}(1,1.5){\psline{->}(0,0)(0.75,0)}
 \rput(1,1.25){\psline{->}(0,0)(0.75,0)}
 \rput{-26.57}(1,1){\psline{->}(0,0)(0.75,0)}

 \rput(4.3,3.25){\small $\Delta n_{a}(E)$}
 \rput(4.3,2.25){\small $+1$}
 \rput(4.3,1.25){\small $0$}
 \rput(4.3,0.25){\small $-1$}

 \rput(6.1,3.25){\small $P(E)$}
 \rput(6.1,2.25){\small $p_{0}^{2}$}
 \rput(6.1,1.25){\small $2p_{0}(1 - p_{0})$}
 \rput(6.1,0.25){\small $(1 - p_{0})^{2}$}

\psline[linestyle=dotted](7.5,0)(7.5,3.75)

 \rput(8.1,3.5){(b)}

 \rput(10.525,3.25){\small $E$}

 \pscircle[fillstyle=solid,fillcolor=black](8.4,1.25){0.25}
 \psline(7.9,1.75)(8.9,0.75) \psline(7.9,0.75)(8.9,1.75)

 \pscircle[fillstyle=solid,fillcolor=black](10.15,2.25){0.25}
 \pscircle[fillstyle=solid,fillcolor=black](10.9,2.25){0.25}

 \pscircle(10.15,1.25){0.25}
 \pscircle[fillstyle=solid,fillcolor=black](10.9,1.25){0.25}

 \pscircle(10.15,0.25){0.25}
 \pscircle(10.9,0.25){0.25}

 \rput{26.57}(8.9,1.5){\psline{->}(0,0)(0.75,0)}
 \rput(8.9,1.25){\psline{->}(0,0)(0.75,0)}
 \rput{-26.57}(8.9,1){\psline{->}(0,0)(0.75,0)}

 \rput(12.2,3.25){\small $\Delta n_{a}(E)$}
 \rput(12.2,2.25){\small $0$}
 \rput(12.2,1.25){\small $+1$}
 \rput(12.2,0.25){\small $+2$}

 \rput(14,3.25){\small $P(E)$}
 \rput(14,2.25){\small $p_{0}^{2}$}
 \rput(14,1.25){\small $2\tilde{p}(1-\tilde{p})$}
 \rput(14,0.25){\small $\tilde{p}^{2}$}

\end{pspicture}
\caption
{Probabilities associated with a reproduction event.  An organism at
  position \locus is shown with an open circle and any other type
  of organism with a solid circle. The columns labelled ``$E$''
  represents the three possible outcomes of a reproduction event; in
  the columns labelled by ``$\Delta n_{a}(E)$'' the variation of
  $n_{a}$ associated with event $E$ is listed. The probabilities
  involved are given in the columns marked $P(E)$, where $p_{0}$
  is the probability of no mutations during a reproduction event and
  $1 - p_{0}$ the probability of at least one mutation while 
  $\tilde{p}$ is defined in Eq.(\ref{eq:ptild}). (a) Reproduction
  originating from \locus. (b) Evaluation of the backflow associated
  with the events $\locusn \neq \locus \rightarrow \locus$.}
\label{fig1}
\end{figure}
The first case happens with probability $P = \rho_{a}(t)\fit^{a}(t)$,
that is, the probability of picking an organism of position \locus, times
the fitness of \locus.  In this event, $n_{a}$ can decrease by one
unit ($\Delta n_{a} = -1$), increase by one unit ($\Delta n_{a} =
+1$), or remain constant ($\Delta n_{a} = 0$), with relative
probabilities as calculated in Fig.~\ref{fig1}(a).

The probability of having $i$ mutations during a single replication is
\begin{equation}
\label{eq:fluxti}
p_{i} = \binom{L}{i}\mut^{i}(1-\mut)^{L - i}, \, \, \forall i = 0,1 
\ldots,L \hspace*{1cm} \mbox{with $\sum_{i = 0}^{L}p_{i} = 1.$}
\end{equation}
From Fig.~\ref{fig1}(a) we can deduce the net contribution to the
population at position \locus by summing over all possible events:
\begin{equation}
  \sum_{E} \Delta n_{a}(E)P(E) = p_{0}^{2} - (1-p_{0})^{2} = 2p_{0} - 1.
\end{equation}

The ``back flow'' contribution occurs with probability
\begin{equation}
  \sum_{b \neq a} \rho_{b}(t)\fit^{b}(t).
\end{equation}

In this case, the variations and the probabilities involved are shown
in Fig.~\ref{fig1}(b). 

In order to mutate from \locus[b] to
\locus, $Ld_{ab}$ mutations are necessary, where
\begin{equation}
\label{eq:Hamming}
  d_{ab} = d(\locus,\locus[b]) = \frac{1}{2L} \sum_{i = 1}^{L} \left| S^{a}_{i} - S^{b}_{i} \right|
\end{equation}
so
\begin{equation}
  \tilde{p} = \mut^{Ld_{ab}}(1 - \mut)^{L(1- d_{ab})}
\label{eq:ptild}
\end{equation}
is the probability of creating an organism in position \locus
originating from position \locus[b].

As $Ld_{ab}$ mutations are needed, the probability involved in a
back-flow contribution from position \locus[b] is, see Fig. \ref{fig1}(b),
\begin{equation}
  \sum_{E} \Delta n_{a}(E)P(E) = 2\tilde{p}(1-\tilde{p})+2\tilde{p}^{2} =
  2\tilde{p} = 2\mut^{Ld_{ab}}(1-\mut)^{L(1-d_{ab})}.
\end{equation}

Thus, the full expression for the difference equation is,
\begin{eqnarray}
  n_{a}(t+1) &=& n_{a}(t) - \rho_{a}(t)\kil  + 
  \rho_{a}(t)\fit^{a}(t)(2p_{0} - 1) \nonumber \\
&& +\, 2\sum_{b \neq a}\rho_{b}(t) \fit^{b}(t) \mut^{Ld_{ab}}(1-\mut)^{L(1-d_{ab})}. 
\label{eq:mastas} 
\end{eqnarray}
This is the equivalent of the quasi-species equation by Eigen {\it et al.}
The main difference is that the rates of production depend on the current
composition in population space.

Summing Eq.(\ref{eq:mastas}) over all positions in genotype space
we find, as expected,
\begin{equation}
  N(t+1) = N(t) - \kil + \langle \fit \rangle.
\end{equation}

From the simulations we know that in the limit of strong interactions
among the individuals, the dynamics is intermittent \cite{chr:tan,hal:agi}.
Extended periods are
dominated by a network of few heavily occupied positions.
These periods, called quasi-Evolutionary Stable Strategies (q-ESS),
are interrupted by sharp bursts where the configuration of the species
change rapidly and significantly.  In order to describe the
dynamics, we impose a stability condition on the
difference equation: we require that within a single
q-ESS, the average number of individuals remains constant.
Moreover, the q-ESS states are dominated by some very fit positions
surrounded by unfit neighbouring positions.
Thus we can neglect the back-flow contribution in the difference equation,
Eq.(\ref{eq:mastas}), and obtain
\begin{equation}
  n_{a}(t+1) = n_{a}(t) + \rho_{a}(t)\left[\fit^{a}(t)(2p_{0}-1)-\kil \right].
\end{equation}

Averaging over time, the equation becomes
\begin{equation}
  \overline{n_{a}} = \overline{n_{a}} +
  \overline{\rho_{a}\fit^{a}}(2p_{0}-1)-\overline{\rho_{a}}\kil .
\end{equation}
Assuming that $\overline{\rho_{a}\fit^{a}} =
\overline{\rho_{a}}\overline{\fit^{a}}$, the
fitness for all positions in the set $\mathcal{S}_{\fit=p_{q}}$:
\begin{equation}
  \label{eq:fqexpr}
  \overline{\fit^{a}} = \frac{\kil}{2p_{0} - 1} \equiv p_{q} .
\end{equation}
With $\mut = 0.008$ we have $p_{0} = (1 - \mut)^{L} = 0.852$ for
$L = 20$; using
$\kil = 0.2$, we find $p_{q} = 0.284$ consistent with the observation
of Fig.~\ref{fig2}.

\begin{figure}[htbp]
  \centering
  \begin{pspicture}(14,7.75)
    \rput(7.5,4.25){\includegraphics[width=12cm,totalheight=7cm]{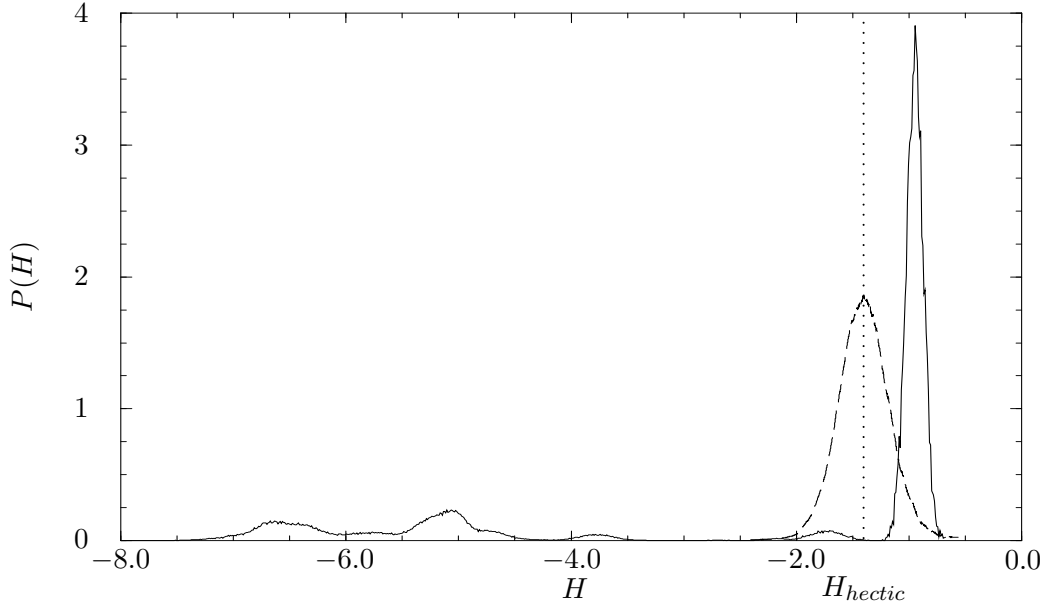}}
    \rput(1.52,0.47){\small $-8.0$}
    \rput(4.5,0.47){\small $-6.0$}
    \rput(7.52,0.47){\small $-4.0$}
    \rput(10.5,0.47){\small $-2.0$}
    \rput(13.52,0.47){\small $0.0$}
    \rput(7.56,0.1){\small $H$}
    \psline[linestyle=dotted](11.4,0.75)(11.4,7.75)
    \rput(11.4,0.1){\small $H_{hectic}$}
    \rput(1,0.75){\small $0$}
    \rput(1,2.5){\small $1$}
    \rput(1,4.25){\small $2$}
    \rput(1,6){\small $3$}
    \rput(1,7.75){\small $4$}
    \rput{90}(0.25,4.25){\small $P(H)$}

  \end{pspicture}
  \caption{The probability density function of the weight function
    $H = \ln \left(\frac{p_{off}}{1-p_{off}}\right)$ during a q-ESS state of a
    simulation (solid line) and during a transition between 2 q-ESS
    states (dashed line).
    During a q-ESS state (solid line) positions
    range in two sets: unfit positions, for which the weight function
    is lower than $-3.0$ and fit positions, for which the fitness is
    greater then the average value $\left< H \right> = \ln \left(
      \frac{1-\kil}{\kil}\right) \approx -1.38 = H_{hectic} $,
    indicated by a vertical dotted line.
    During a transition (dashed
    line) the fitness of all positions is normally distributed around
    $H_{hectic}$ where all positions reproduce (on average) at the same
    rate, equal to the killing rate.
 Notice the support of the
weight function in the hectic phase exceeds $H_q$, ensuring that
positions in genotype space are able to fulfill the q-ESS balance
Eq.(\ref{eq:eqval}).
The parameters (for precise definitions, see Refs.
\cite{chr:tan, hal:agi}) are $\kil = 0.2$, $\mu = 1/1000 \cdot \ln \left(
    \frac{1-\kil}{\kil} \right) \approx 0.0014, C = 10.0 \ \mbox{and}
    \ \mut = 0.008$.
}
  \label{fig2}
\end{figure}

Neglecting the back flow is valid if all terms 
\[
\rho_{b}(t)\fit^{b}(t) \mut^{Ld_{ab}} \left[ 1-\mut \right] ^{L(1-d_{ab})} = 
\rho_{b}(t)\fit^{b}(t) \mut^{Ld_{ab}} \left[1 - L(1-d_{ab})\mut +
  \cdots \right]
\] 
are small. Since $\mut \ll 1$, the leading term is $\rho_{b}(t)
\fit^{b}(t) \mut^{Ld_{ab}}$.  This can be neglected if $Ld_{ab} > 1$.
With $Ld_{ab} = 1$ it can be neglected since none of the nearest neighbours
are fit, as $\fit^{b}(t) \ll 1$.

\section{The error threshold}
\label{sec:error}
The discussion of the q-ESS state was made with the implicit
assumption of the existence of q-ESS states.  We will find here that
we can establish qualitative arguments that ensure the existence of
the q-ESS states.

We have seen that q-ESS states are possible only if the interactions
are important in the weight function.  Furthermore, the average fitness
$p_q$ of the fit positions in the q-ESS state is given by
\begin{equation}
\label{eq:eqval}
  p_{q} = \frac{\kil}{2(1-\mut)^L - 1}
\end{equation}
and thus is related to the mutation rate.
This relation states that the fit positions are those that are able to
counterbalance the killing by the production of offspring.

Equation (\ref{eq:eqval}) is the starting point for determining a
necessary condition for the existence of a q-ESS state.  We have
investigated the behaviour of the dynamics as a function of mutation
rate. The results are illustrated in Fig.~\ref{fig3}.

\begin{figure}[htbp]
  \centering
  \begin{pspicture}(14,17)
    \rput[t](7.5,17){\includegraphics[width=13cm,totalheight=16cm]{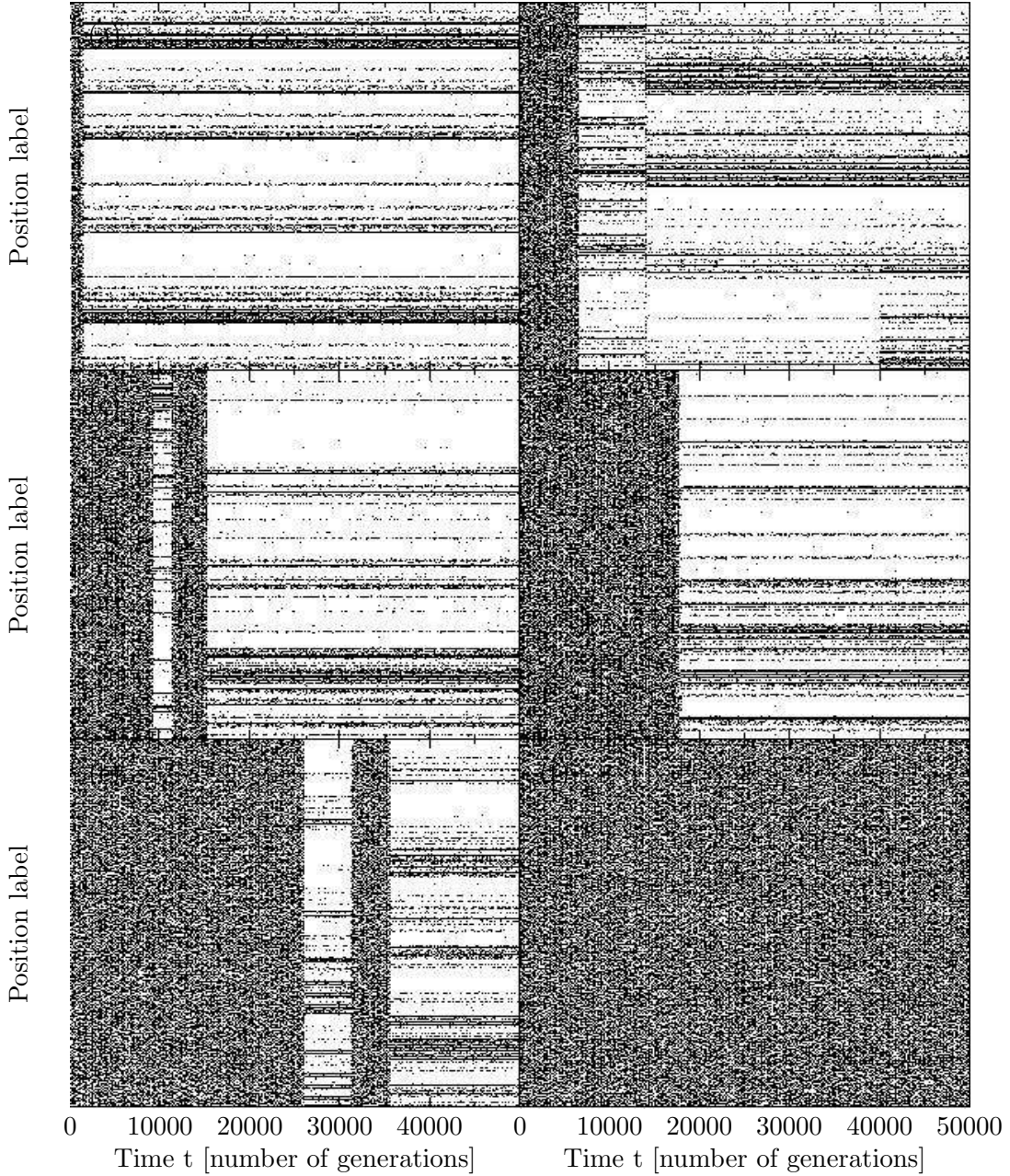}}
    \rput{90}(0.25,3.667){\small Position label}
    \rput{90}(0.25,9){\small Position label}
    \rput{90}(0.25,14.333){\small Position label}
    \rput(4.25,0.25){\small Time t [number of generations]}
    \rput(10.75,0.25){\small Time t [number of generations]}

    \rput(1,0.75){\small $0$}
    \rput(2.3,0.75){\small $10000$}
    \rput(3.6,0.75){\small $20000$}
    \rput(4.9,0.75){\small $30000$}
    \rput(6.2,0.75){\small $40000$}

    \rput(7.5,0.75){\small $0$}
    \rput(8.8,0.75){\small $10000$}
    \rput(10.1,0.75){\small $20000$}
    \rput(11.4,0.75){\small $30000$}
    \rput(12.7,0.75){\small $40000$}
    \rput(14,0.75){\small $50000$}

    \rput(1.5,16.5){\small (a)}
    \rput(8,16.5){\small (b)}
    \rput(1.5,11.167){\small (c)}
    \rput(8,11.167){\small (d)}
    \rput(1.5,5.833){\small (e)}
    \rput(8,5.833){\small (f)}
  \end{pspicture}
  \caption[Moving towards the error threshold]{Occupation plots for
    different values of the mutation rate. The $y$-axis refers to an
    arbitrary ennumeration of all positions in genotype space. Occupied
    positions are indicated by a black dot. Results shown are for $\kil
    = 0.2$, $\mu = 1/1000 \cdot \ln \left( \frac{1-\kil}{\kil}
    \right)$ and $C = 0.05$). (a) Mutation rate: $\mut = 0.009$.  The
    initial transient is extended.  (b) Mutation rate: $\mut =
    0.00925$.  The initial transient has the same extension of any
    q-ESS state. (c) Mutation rate: $\mut = 0.0095$.  The transition
    between two q-ESS state are extended.  (d) Mutation rate: $\mut =
    0.01$.  The initial transient is very extended. (e) Mutation rate
    $\mut = 0.0104$.  The initial transient and any transitions are
    extensively hectic. (f) Mutation rate $\mut = 0.0108$.  There is
    no q-ESS state.}
  \label{fig3}
\end{figure}

For increasing \mut, the duration of q-ESS states decreases.  Above a
threshold \ert of the mutation rate \mut, there are no
more q-ESS states: the dynamics is completely hectic.
For intermediate values of \mut, the transitions between two
q-ESS states are extended and the initial transient can be very long.

This numerical result shows that the model defines an error
threshold for the mutation rate above which no q-ESS state exists.

From Eq.(\ref{eq:eqval}) we obtain for the weight function
\begin{equation}\label{eq:Hq}
H_{q} = \ln \left( \frac{p_{q}}{1-p_{q}} \right) = \ln \left(
  \frac{\kil}{2p_{0}-1-\kil} \right).
\end{equation}

When the mutation rate is close to \ert, most of the simulations are
in hectic states, for which the fitness is equal to \kil and therefore
we might assume that the weight function is equal to
\begin{equation}\label{eq:Hh}
H_{hectic} = \ln \left( \frac{\kil}{1-\kil} \right) .
\end{equation}

Stable q-ESS states can only develop from a hectic phase when some positions,
due to fluctuations, acquire sufficient fitness to be consistent with the
q-ESS balance given by Eq.(\ref{eq:eqval}). That is, fluctuations in the
weight functions in the hectic phase must allow
\begin{equation}\label{eq:inico}
H_{hectic} + \frac{\alpha}{C} \geq H_{q}
\end{equation} 
where $\alpha \in (0,1)$ describes the width of the 
distribution of weight functions in the hectic phase, see
Fig. \ref{fig2} and $C$ determines the width of the distribution of the
possible coupling strengths between the individuals. Small $C$
corresponds to the strong coupling regime while large $C$ corresponds
to the weak coupling limit.
Using Eq.(\ref{eq:Hq}) and Eq.(\ref{eq:Hh}) we obtain
\begin{equation}
  \ln \left( \frac{\kil}{1-\kil} \right) + \frac{\alpha}{C} \geq
  \ln \left( \frac{\kil}{2p_{0}-1-\kil} \right) 
\end{equation}
which, translated to the mutation rate \mut, becomes
\begin{equation}\label{eq:ert}
  \mut  \leq 1 - \left[ \frac{e^{-\alpha/C}(1-\kil)+1+\kil}{2}
  \right]^{1/L} = \ert
\end{equation}
Eq.(\ref{eq:ert}) defines the functional dependency of the error
threshold in terms of $\alpha$, $C$ and \kil. In Fig.~\ref{fig4} we
use $\alpha$ as a fitting parameter and show \ert as a function of $C$.
\begin{figure}[htbp] \centering
  \begin{pspicture}(14,8)
    \rput[bl](1.5,0.85){\includegraphics[width=12cm]{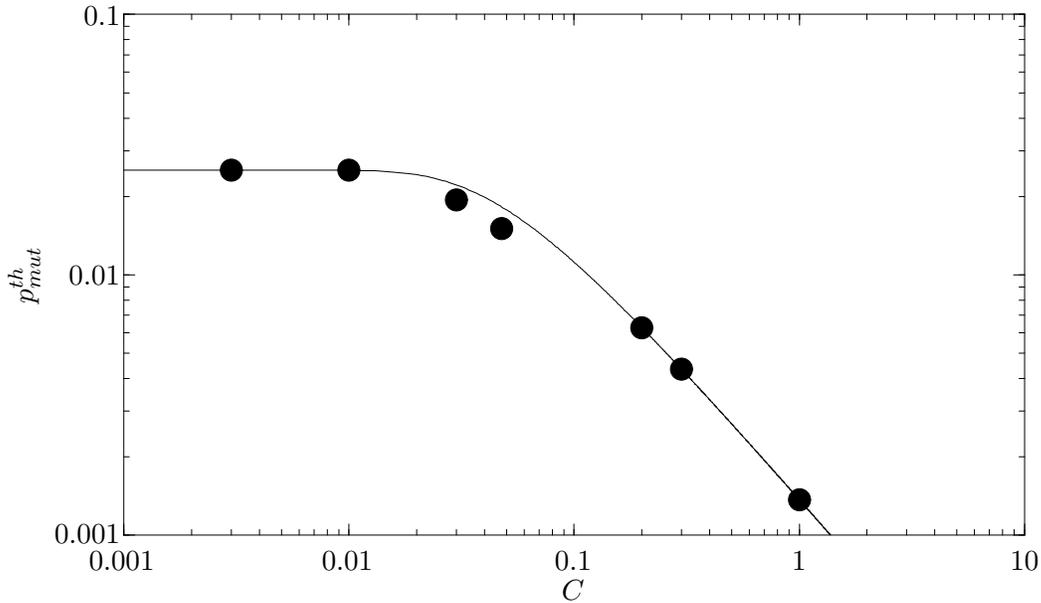}}
    \rput(1.5,0.5){\small $0.001$}
    \rput(4.5,0.5){\small $0.01$}
    \rput(7.5,0.5){\small $0.1$}
    \rput(10.5,0.5){\small $1$}
    \rput(13.5,0.5){\small $10$}
    \rput(7.5,0.1){\small $C$}
    \rput[r](1.48,0.85){\small $0.001$}
    \rput[r](1.48,4.3){\small $0.01$}
    \rput[r](1.48,7.75){\small $0.1$}
    \rput{90}(0.2,4.3){\small \ert}
  \end{pspicture}
  \caption[The error threshold]{The computational determination of the 
    error threshold. The loss of q-ESS states occurs for mutation rates
    above the solid circles. The data, compared with the theoretically
    predicted error threshold \ert (solid line), indicate a value of
    $\alpha = 0.07$, see Eq.(\ref{eq:ert}). The 
    parameters of the simulations are $L = 20, \mu = 0.005$ and
    \kil = 0.2.}\label{fig4}
\end{figure}

The error
threshold has been determined numerically by iterating many simulations with
increasing value of the mutation rate for a given $C$. When no
q-ESS emerges, we have reached the error threshold; the lowest \mut
for which only a hectic states exists is the estimated value of \ert.

The numerical results confirm the theoretical predictions given by
Eq.(\ref{eq:ert}) and, qualitatively, are in line with the results of
Eigen \textit{et al.} \cite{eig:sel,eig:qua}.
The transition in the Tangled Nature model appears to be
sharp, that is, for values of \mut greater than \ert q-ESS
states are impossible, while for $\mut \leq \ert$ q-ESS are
possible, see Fig~\ref{fig3}.

Since the factor $\alpha$ represents the width of the distribution
of the weight function during a hectic state
it is linked to 
$\mathbf{J}=\left\{ J_{ab} \right\}$, the set of interactions, and
also to $\mu$. This makes it difficult to analytically determine
$\alpha$.

\section{Discussion}
\label{diss}

In the Tangled Nature model the
competition of the organisms is described by the mutual
interactions, creating a dynamical rugged fitness landscape where the
fitness of a position is determined by the temporal evolution of
the ecology.  The dynamics, illustrated in~\cite{chr:tan,hal:agi} selects few
heavily occupied positions in genotype space surrounded by other sequences in
the immediate vicinity.  The central positions are the only
able to reproduce actively. They sustain themselves and all the
surrounding ecology.
This situation is possible only as long as the mutual interactions,
are sufficient to counterbalance the dispersive
action caused by mutations.

Thus we have derived an interpretation of the Tangled Nature model as an
evolutionary quasi-species model.
In the Tangled Nature model however, the fitness landscape is not
fixed. Due to the frequency dependent fitness landscape, the
Tangled Nature model allows the emergence of multiple co-existing quasi-species
during q-ESS states. Also, it should be noted, that in contrast to the
model by Eigen {\it et al.} \cite{eig:sel,eig:qua}}, the quasi-species in the
Tangled Nature model are not absolute quantities but may change from
one q-ESS to another.

We have discussed and identified the error threshold in the Tangle Nature
model as the mutation rate at which the model is unable to support,
over extended periods in time, the occupation of well defined multiple
co-existing genotypes. A formula for the parameter dependence of the error
threshold was derived, see Eq.(\ref{eq:ert}). In particular, the error threshold
depends on genome length as $1/L$, (for large L) which is consistent with the
findings in the quasi-species models, see Refs. \cite{dro:bio,gal:dif}. This
result suggests that the mutation rate per base pair itself is
subject to selection in a way to make the mutation per base pair decrease
with increasing genome length. This is indeed observed in nature.

We are extremely grateful to Matt Hall for very helpful
discussions.
K. C. gratefully acknowledges
the financial support of U.K. EPSRC through Grant No. GR/R44683/01.

\bibliography{biblio}

\end{document}